\documentclass[12pt]{article}
\usepackage[latin1]{inputenc} %paquete para las tildes
\usepackage{amssymb}
\usepackage{amsfonts} %paquete para escribir letras como el cuerpo de los reales
\usepackage{amsthm} %paquete matemÃ¡tico
\usepackage{amsmath} %paquete matemÃ¡tico
\usepackage{hyperref}
\usepackage{color}
\usepackage{young}

\def\ni{\noindent}
\def\nn{\nonumber}

\def \bc {\begin{center}}
\def \ec {\end{center}}
\def \bi {\begin{itemize}}
\def \ei {\end{itemize}}

\def \ba {\begin{array}}
\def \ea {\end{array}}

\def \bea {\begin{eqnarray}}
\def \eea {\end{eqnarray}}

\def \be {\begin{equation}}
\def \ee {\end{equation}}

\newcommand{\la}{\langle}
\newcommand{\ra}{\rangle}

\def \um {\frac{1}{2}}

\def\tr {\mathrm{tr}}

\def\mbN {{\mathbb N}}

\def\mbS {{\mathbb S}}
\def\mbC {{\mathbb C}}

 \def\cD {{\cal D}}
 \def\cN {{\cal N}}

\newtheorem{thm}{Theorem}[section]
\newtheorem{cor}[thm]{Corollary}
\newtheorem{lem}[thm]{Lemma}
\newtheorem{prop}[thm]{Proposition}

\theoremstyle{remark}

\begin{document}

\begin{center}
{\Large {\bf Coherent states on the Grassmannian   $U(4)/U(2)^2$:
Oscillator realization and bilayer fractional quantum Hall systems}}
\end{center}
\bigskip
%\bigskip

\centerline{{\sc M. Calixto}\footnote{Corresponding author: calixto@ugr.es}
 and {\sc E. Pérez-Romero}}

\bigskip

\bc {\it  Departamento de Matemática Aplicada,
Facultad de Ciencias, Campus de Fuentenueva,  18071 Granada, Spain}
\ec

\bigskip
\begin{center}
{\bf Abstract}
\end{center}
\small
\begin{list}{}{\setlength{\leftmargin}{3pc}\setlength{\rightmargin}{3pc}}
\item
Bilayer quantum Hall (BLQH) systems, which underlie a $U(4)$ symmetry, display unique quantum coherence effects. 
We study coherent states (CS) on the complex Grassmannian $\mathbb G_2^4=U(4)/U(2)^2$, orthonormal 
basis, $U(4)$ generators and their matrix elements in the reproducing kernel Hilbert space 
$\mathcal H_\lambda(\mathbb G_2^4)$ of analytic square-integrable 
holomorphic functions on $\mathbb G_2^4$, which carries a unitary irreducible 
representation of $U(4)$ with index $\lambda\in\mathbb N$. A many-body representation of the previous construction is introduced 
through an oscillator realization of the $U(4)$ Lie algebra generators in terms of eight boson operators. 
This particle picture allows us for a physical interpretation of our abstract mathematical construction in the 
BLQH jargon. In particular, the index $\lambda$ is related to the number of flux quanta bound to a bi-fermion 
in the \emph{composite fermion} picture of Jain for fractions of the filling factor $\nu=2$. The simpler, and better 
known, case of spin-$s$ CS on the Riemann-Bloch sphere $\mathbb{S}^2=U(2)/U(1)^2$ is also treated in parallel, of which 
Grassmannian $\mathbb G_2^4$-CS can be regarded as a generalized (matrix) version.

\end{list}
\normalsize %\setlength{\baselineskip}{14pt}

%Phase transitions: quantum Hall effects, 73.43.Nq
%Tunneling: in quantum Hall effects, 73.43.Jn



%\newpage
\section{Introduction}

Since Schr\"odinger first introduced in 1926 the notion of (canonical) Coherent States (CS) of the harmonic oscillator, 
the subject of CS has grown and permeates almost all branches of quantum physics (see e.g. \cite{Klauder} and \cite{Vourdas,JPA} 
for  old and recent reviews). Besides, some other important topics in applied mathematics, like the theory of wavelets, 
are also related to the notion of CS \cite{Gazeau}. Later in 1972, Gilmore \cite{Gilmore1,Gilmore11} and Perelomov \cite{Perelomov,Perelomovbook} realized that canonical CS
were rooted in group theory (the Heisenberg-Weyl group)  and generalized the concept for other type of groups. Actually, 
Gilmore introduced an algorithm \cite{Gilmore2}, which makes use of CS as variational states to approximate the ground state energy, 
to study the classical, thermodynamic or mean-field, limit of some algebraic quantum models. This algorithm has proved to be specially suitable to 
analyze the phase diagram of Hamiltonian models undergoing a quantum phase transition. 

Among all physical models where CS play a relevant role, we want to highlight the Quantum Hall Effect (QHE). 
Several interesting text books on the subject are namely \cite{Prange,EzawaBook,Jainbook,Jacak}. We briefly 
remind that QHE deals with 
two-dimensional electron systems subjected to a perpendicular magnetic field $B$. Electrons 
make cyclotron motions and 
their energies are quantized into Landau levels. The number density of magnetic flux quanta is $\rho_\phi=B/\phi$, 
where $\phi=2\pi\hbar/e$ is the flux unit. One electron occupies an area $2\pi\ell_B$  with $\ell_B=\sqrt{\hbar c/eB}$ 
the magnetic length and the \emph{filling factor} is $\nu=\rho_0/\rho_\phi$ with $\rho_0$ the electron number density. 
QHE has attracted renewed attention owing to its peculiar features associated with
quantum coherence. In fact, bilayer quantum Hall (BLQH) systems are much more interesting because they exhibit 
unique effects originating in the interlayer interaction, like the development of spontaneous quantum coherence across the layers.  
A bilayer system is made by trapping electrons in two thin layers at the interface of semiconductors.
Electrons are transferable between the two layers by applying bias voltages. 
In the BLQH  system one Landau site may
accommodate four isospin states $|b\uparrow\rangle, |b\downarrow\rangle, |a\uparrow\rangle$ and $|a\downarrow\rangle$ in
the lowest Landau level,  where $|b\uparrow\rangle$ (resp. $|a\downarrow\rangle$) means  that the
electron is in the bottom layer ``$b$'' (resp. top  layer ``$a$'') and its spin is up (resp. down), and so on.
Therefore, the $U(4)$ symmetry underlies the BLQH system provided the cyclotron energy is large enough.
The driving force of quantum coherence is the Coulomb exchange interaction, which is described by an anisotropic $SU(4)$
nonlinear $\sigma$-model in BLQH systems \cite{EzawaBook}. Actually, it is the interlayer exchange interaction which 
develops the interlayer coherence. The lightest topological charged excitation in the BLQH system is a (complex projective) 
$\mathbb CP^3=U(4)/[U(1)\times U(3)]$ skyrmion for filling factor $\nu=1$ and a (Grassmannian) 
$\mathbb G_2^4=U(4)/[U(2)\times U(2)]$ bi-skyrmion (two $\mathbb CP^3$ skyrmions carrying total charge $2e$)  
for filling factor $\nu=2$. The Coulomb exchange interaction for this last case is described by a Grassmannian
$\mathbb G_2^4$  $\sigma$-model and the dynamical field  is a  Grassmannian field  
$Z=z^\mu\sigma_\mu$ \cite{Ezawabisky} [$\sigma_\mu$ are the Pauli matrices in
\eqref{Pauli}] carrying four complex field degrees of freedom $z^\mu\in\mathbb C$, $\mu=0,1,2,3$. Also, the parameter 
space characterizing the $SU(4)$-invariant ground state in the BLQH system at $\nu=2$ is precisely $\mathbb G_2^4$ \cite{EzawaPRB}.

Just to mention that other construction of coherent states on the Grassmannian $\mathbb G^N_2=U(N)/ [U(2)U(N-2)]$ 
(space of complex two planes in $\mathbb C^N$) has been recently discussed in \cite{loopQG}, but in connection 
with loop quantum Gravity, where the quantum states of geometry are the so-called spin network states.

In this article we make a quite thorough (mathematical) study of CS on $\mathbb G_2^4$, which we are sure that will 
be of great physical utility as variational states to study the semi-classical (and thermodynamical limit) analysis 
of the BLQH system and its quantum phase transitions, just like standard spin-$s$ CS are essential for semi-classical 
studies of quantum phase transitions in boson condensates. Firstly we follow a geometric 
approach to the construction of CS on $\mathbb G_2^4$, in part inspired by the method of orbits in geometric quantization  
due to Kirillov-Kostant-Souriau \cite{Kirillov,Kostant,Souriau} and the Borel-Weil-Bott theorem \cite{FultonHarris}, 
which relate  quantization, geometry and the representation theory
for classical groups. In order to connect this abstract construction with the ``many body picture'', we introduce 
an oscillator realization of the $u(4)$ Lie algebra in terms of eight boson creation, $a_\mu^\dag, b_\mu^\dag$, and annihilation,
$a_\mu, b_\mu, \mu=0,1,2,3$, operators. This realization differs from the standard Schwinger boson representation of $u(4)$ in terms of 
four  bosons, leading to the totally symmetric representation and related to the  Grassmannian $\mathbb G^4_1=\mathbb CP^3$. 
A similar oscillator realization to ours, but for the (non-compact) pseudo-Grassmannian $U(2,2)/U(2)$,
has been recently considered in \cite{Grosse}, in the context of deformation quantization, 
recovering some old results of R\"uhl \cite{Ruhl0,Ruhl1} concerning CS on the conformal group (see also \cite{EMSMTA,spinning,unruh} 
on this subject). Other boson realizations of the $u(N)$ Lie algebra appear in the literature, namely by 
Moshinsky  \cite{MoshinskyPL,MoshinskyNPB,MoshinskyJMP,MoshinskyBook} in the context of nuclear physics, 
who demonstrated that the irreps  of a unitary algebra are characterized by a partition of the number of particles
involved and he showed that a basis of the space underlying the irrep can be constructed
from the so-called ``highest-weight polynomial". Coherent states and oscillator realizations 
for $SU(N)$ have also been discussed in \cite{Mathur}, and an identification and state labeling of the class of 
irreps of $SU(4)$ with respect  to $S(U(2)\times U(2))$ have been  identified in \cite{MacfarlaneJPA} 
(see also \cite{Brunet,Quesne}). However, we do not find a clear connection with our construction, which is 
specially designed to the study of BLQH systems.

The paper is organized as follows. In Section \ref{su4coord} we remind the Lie algebra structure and coordinate systems 
of $U(4)$ adapted to the fibration $U(2)^2\to U(4)\to \mathbb G_2$  (since there is no confusion, from now on we shall use
the short-hand $\mathbb G_2=\mathbb G_2^4$). In Section \ref{sec2} we construct a 
CS system labeled by points of $\mathbb G_2$ in the (reproducing kernel) Hilbert space ${\cal H}_\lambda(\mathbb G_2)$ 
of analytic square-integrable holomorphic functions on $\mathbb G_2$ with a given measure (orthonormality relations 
are proved in the Appendix \ref{appendixortho}). This corresponds to a given  
square-integrable irreducible representation of $U(4)$ with positive integer index $\lambda$, and we identify the Young tableau 
associated with it, which motivates the ``particle picture'' construction later in Section \ref{oscisec} (those readers more 
acquainted with the many-body picture might skip Section \ref{sec2} in a first reading and go to Section \ref{oscisec}). 
Before, in Section \ref{sec3} we explicitly compute the generators (pseudospin ladder, imbalance, angular momentum, etc, operators) 
of the representation of $U(4)$ on ${\cal H}_\lambda(\mathbb G_2)$ 
and their matrix elements in an orthonormal basis. In Section \ref{oscisec}, we introduce an oscillator realization of the 
$u(4)$ Lie algebra in terms of eight boson operators, and express the orthonormal basis of ${\cal H}_\lambda(\mathbb G_2)$ 
in terms of the Fock basis with constraints in the occupancy numbers.  An expression of Grassmannian CS as 
Bose-Einstein-like condensates is also provided. The spin-frozen case, which is described by standard pseudospin-$s$ CS on the Riemann-Bloch sphere 
$\mathbb{S}^2=U(2)/U(1)^2$, is treated in parallel all along the paper, to better appreciate the role played by spin in BLQH systems and to 
stress the similarities and differences between  $\mathbb G_2$ and $\mathbb S^2$ CS, the first  being a generalized (matrix) version of the second ones. 
Section \ref{comments} is devoted to some comments on the (flux quanta) physical meaning of the representation index $\lambda$ 
and its relation with the \emph{composite fermion} picture of Jain \cite{JainPRL,Jainbook} in the fractional quantum Hall effect.

\section{\label{su4coord}The group $U(4)$: coordinate systems and generators}

Let us firstly describe very briefly the structure of the group $U(4)$ of unitary $4\times 4$ matrices, reminding its Lie algebra basis and
putting coordinates on it. In this article we are interested
in the Lie algebra basis adapted to the noncanonical chain of subgroups
\[U(4)\supset U(2)\times U(2)\supset U(1)\times U(1).\]
The corresponding matrix representation is useful, for instance, when studying isospin $SU(4)$ symmetry in
bilayer spin (namely, quantum Hall) systems, to emphasize
the spin $SU(2)$ symmetry in the, let us say,  bottom ($b$ or pseudospin $-1/2$) and top ($a$ or pseudospin $1/2$) layers.
The pseudospin rotates when particles are transfered from one layer to the other. More precisely, we denote
the $U(4)$ generators in the fundamental representation by $\tau_{\mu\nu}\equiv\sigma_\mu\otimes\sigma_\nu, \, \mu,\nu=0,1,2,3$ where
\be \sigma_0=\left(\ba{cc} 1& 0
\\ 0 &1\ea\right),\;\sigma_1=\left(\ba{cc} 0& 1
\\ 1 &0\ea\right),\;\sigma_2=\left(\ba{cc} 0& -i
\\ i &0\ea\right),\;\sigma_3=\left(\ba{cc} 1& 0
\\ 0 &-1\ea\right).\label{Pauli}\ee
denote the Pauli matrices (plus $\sigma_0$). We shall introduce, for convenience, the interlayer ladder
matrices
\be
\tau_{+\mu}\equiv\um(\tau_{1\mu}+ i \tau_{2\mu})=\left(\ba{cc} 0& \sigma_\mu \\ 0
&0\ea\right)\,,\;\; \tau_{-\mu}\equiv\um(\tau_{1\mu}- i \tau_{2\mu})=\left(\ba{cc} 0& 0
\\ {\sigma}_\mu &0\ea\right)\label{taupm}
\ee
and the Lorentz-like generators
\be
\mathfrak{m}_{\mu\nu}=\frac{1}{4}\left(\ba{cc} \sigma_\mu\check{\sigma}_\nu-\sigma_\nu\check{\sigma}_\mu & 0\\
0&\check{\sigma}_\mu\sigma_\nu-\check{\sigma}_\nu\sigma_\mu\ea\right),\label{Mmunu}
\ee
where $\check{\sigma}_\nu\equiv \sigma^\nu=\eta^{\nu\mu}\sigma_\mu$ and we shall use the metric $\eta_{\mu\nu}=\mathrm{diag}(1,-1,-1,-1)$ to rise and lower indices. The
Einstein summation convention will also be used unless otherwise stated\footnote{Although we are in principle in a 
non-relativistic setting, relativistic notation turns out to be quite convenient.}. Note that $\mathfrak{m}_{\mu\nu}$ can be expressed in terms of $\tau_{0j}$ and
$\tau_{3j}, j=1,2,3$, as: $\mathfrak{m}_{0j}=-\um \tau_{3j}$ and $\mathfrak{m}_{jk}=\frac{-i}{2}\epsilon_{jkl}\tau_{0l}$, 
with $\epsilon_{jkl}$ the Levi-Civita symbol. The 
$su(4)$ commutation relations are written in terms of $\tau_{\pm\mu}, \mathfrak{m}_{\mu\nu}$ and the ``pseudospin 
third component'' $\tau_{30}$ as (we denote 
$\check\tau_{\pm\mu}=\tau_{\pm}^\mu=\eta^{\mu\nu}\tau_{\pm\nu}$):
\begin{eqnarray}
[\mathfrak{m}_{\mu\nu},\mathfrak{m}_{\alpha\beta}]=\eta_{\nu\alpha}\mathfrak{m}_{\mu\beta}+\eta_{\mu\beta}\mathfrak{m}_{\nu\alpha}
-\eta_{\mu\alpha}\mathfrak{m}_{\nu\beta}-\eta_{\nu\beta}\mathfrak{m}_{\mu\alpha},\nn\\ 
{}[\check\tau_{-\mu},\mathfrak{m}_{\alpha\beta}]=\eta_{\mu\alpha}\check\tau_{-\beta}-\eta_{\mu\beta}\check\tau_{-\alpha},\;\;
[\check\tau_{+\mu},\mathfrak{m}_{\alpha\beta}]=\delta_{\mu\alpha}\tau_{+\beta}-\delta_{\mu\beta}\tau_{+\alpha},\nn\\
{}[\tau_{+\mu},\check\tau_{-\nu}]= \eta_{\mu\nu}\tau_{30}+2\mathfrak{m}_{\mu\nu},\;\; 
{}[\tau_{30},\tau_{\pm\mu}]=\pm 2\tau_{\pm\mu},\nn\\ {}[\tau_{30},\mathfrak{m}_{\mu\nu}]=0, [\tau_{\pm\mu},\tau_{\pm\nu}]=0.
\label{su4commutators}
\end{eqnarray}
The linear Casimir operator is
$C_1=\tau_{00}$. The quadratic Casimir operator can be written in several forms as
\bea
 C_2&=&\frac{1}{4}\delta^{\mu\nu}\delta^{\alpha\beta}\tau_{\mu\alpha}\tau_{\nu\beta}-\frac{1}{4}\tau^2_{00}
 \nn\\
 &=&\frac{1}{4}(\tau_{0\mu}
 \check{\tau}_0^\mu+ \tau_{3\mu}\check{\tau}_3^\mu)+\um(\check\tau_{-\mu}
 {\tau}_+^\mu+\check{\tau}_{+\mu} \tau_{-}^\mu)-\frac{1}{4}\tau_{00}^2\nn\\
 &=& \frac{1}{4}\tau_{30}^2+ 2(\vec{s}_a^2+\vec{s}_b^2)+
 \um(\check\tau_{-\mu}
 {\tau}_+^\mu+\check{\tau}_{+\mu} \tau_{-}^\mu)
,\label{Casimir}
\eea
which, for the current fundamental (four-dimensional) representation, is simply $\frac{15}{4}\tau_{00}$. 
In the last equality we have also introduced the angular momentum
\be
{s}_{aj}=\frac{1}{4}(\check{\tau}_{0j}+\check{\tau}_{3j})=\left(\ba{cc}  -\um\sigma_j &0 \\ 0
&0\ea\right),\; {s}_{bj}=\frac{1}{4}({\tau}_{0j}-{\tau}_{3j})=\left(\ba{cc}  0 &0 \\ 0
& \um\sigma_j\ea\right), j=1,2,3,\label{angmom}
\ee
of the top ($a$) and bottom ($b$) layers.
The relative sign between $\vec{s}_a$ and $\vec{s}_b$ has a sense that will be explained
later (it could be assimilated to the space-fixed and body-fixed rigid-rotor angular momentum operators).
Note that $\vec{s}_a^2+\vec{s}_b^2= -\frac{1}{4}\mathfrak{m}_{\mu\nu}\mathfrak{m}^{\mu\nu}$. In the BLQH literature
\cite{EzawaBook} it is customary to define the spin $\tau^{\mathrm{spin}}_j=\tau_{0j}$ and
pseudospin $\tau^{\mathrm{ppin}}_j=\tau_{j0}$ matrices, together with the remaining 9 isospin matrices $\tau_{jk}$.
Note that $\tau^{\mathrm{spin}}_j=2(s_{bj}-s_{aj})$.

The fundamental representation of the group $U(4)$ is defined as usual
\be U(4)=\left\{g=\left(\ba{cc} A& B
\\ C &D\ea\right)\in {\rm Mat}_{4\times 4}(\mathbb C):  g^\dag  g=1=g g^\dag \right\},\label{su4} \ee
where, in terms of the $2\times 2$
complex matrices $A,B,C,D$ in (\ref{su4}), the restrictions are explicitly written as
\be g^\dag g=1\Leftrightarrow \left\{\ba{r} D^\dag
D+B^\dag B=\sigma_0\\ A^\dag A+C^\dag C=\sigma_0\\ A^\dag B+C^\dag
D=0,\ea\right.\label{mim}\ee
together with those restrictions of $gg^\dag=1$. In this article we
shall use a set of complex coordinates to parametrize $U(4)$. This parametrization will be adapted
to the complex Grassmannian $\mathbb G_{2}=U(4)/U(2)^2$. It can be obtained through a
block-orthonormalization process of
the matrix columns of:
\be \left(\ba{cc} \sigma_0& 0
\\ -Z^\dag &\sigma_0\ea\right)\rightarrow g=\left(\ba{cc} \sigma_0& Z
\\ -Z^\dag &\sigma_0\ea\right)\left(\ba{cc} \Delta_1& 0
\\ 0 &\Delta_2\ea\right), \left\{ \ba{l} \Delta_1=(\sigma_0+ZZ^\dag)^{-1/2}\\
\\ \Delta_2=(\sigma_0+Z^\dag Z)^{-1/2}\ea\right..
\nn\ee
Actually, we can identify
\begin{eqnarray}
 &Z=Z(g)=BD^{-1}=-A^{\dag-1}C^\dag,\;\; Z^\dag=Z^\dag(g)=-CA^{-1}=D^{\dag-1}B^\dag,& \label{zzdag}\\
&\Delta_1=(AA^\dag)^{1/2},\Delta_2=(DD^\dag)^{1/2}\label{zeta}.&\nn
\end{eqnarray}
The positive-matrix conditions
$AA^\dag>0$ and $DD^\dag>0$ are then equivalent to:
\be \sigma_0+ZZ^\dag>0,\; \sigma_0+Z^\dag Z>0.\label{positive}\ee

Let us conclude this section by giving a complete local parametrization of $U(4)$ adapted
to the fibration $U(2)^2\to U(4)\to \mathbb G_2$. Any element $g\in U(4)$ (in the
present patch, containing the identity element) admits the Iwasawa
decomposition
\be g=\left(\ba{cc} A& B
\\ C &D\ea\right)=\left(\ba{cc} \Delta_1& Z\Delta_2
\\ -Z^\dag\Delta_1 &\Delta_2\ea\right)\left(\ba{cc} U_1& 0
\\ 0 &U_2\ea\right),\label{Iwasawa}\ee
where the matrices
\be U_1=\Delta_1^{-1}A,\; U_2=\Delta_2^{-1}D \nn\ee
belong to $U(2)$ and represent spin rotations in the top and bottom layers, respectively. Likewise, a
parametrization of any $U\in U(2)$ (in a patch containing the
identity), adapted to the quotient $\mathbb CP^1=\mathbb S^2=U(2)/U(1)^2$ (the Hopf fibration) is

\be U=\left(\ba{cc} \mathrm a&  \mathrm b
\\  \mathrm c &  \mathrm d\ea\right)=\left(\ba{cc} \delta & z\delta
\\ -\bar{z}\delta & \delta\ea\right)\left(\ba{cc} u_1& 0
\\ 0 & u_2\ea\right),\label{Iwasawa2}\ee
where $z= \mathrm b/ \mathrm d\in \overline{\mathbb C}\simeq \mathbb S^2$ (the
one-point compactification of $\mathbb C$ by inverse stereographic
projection), $\delta=(1+z\bar{z})^{-1/2}$ and the phases $u_1= \mathrm a/| \mathrm a|,
u_2= \mathrm d/| \mathrm d|$.

\section{Coherent states, closure relations and orthonormal basis}\label{sec2}

Firstly, let us consider the Hilbert space $L^2(U(4),d\mu)$ of square integrable complex
functions $\psi(g)$ on $U(4)$ with  invariant
scalar product
\be\la \psi|\psi'\ra=\int_{U(4)}
d\mu(g){\psi(g)}\overline{\psi'(g)}\label{scalarprod}\ee
given through the invariant Haar measure $d\mu(g)$, which can be decomposed as:
\be\ba{rcl}
d\mu(g)&=&
\left.d\mu(g)\right|_{\mathbb G_2}\left.d\mu(g)\right|_{U(2)^2},\\
\left.d\mu(g)\right|_{\mathbb G_2}&=&
\det(\sigma_0+Z^\dag Z)^{-4}|dZ|,\\
\left.d\mu(g)\right|_{U(2)^2}&=& dv(U_1)dv(U_2),\ea\label{Haarmeasure}\ee
where we are denoting by $dv(U)$ the Haar measure on $U(2)$,
which can be in turn decomposed as:
\bea
dv(U)&=& \left.dv(U)\right|_{\mathbb S^2} \left.dv(U)\right|_{U(1)^2},\nn\\
\left.dv(U)\right|_{\mathbb S^2}
 &=& (1+z\bar
z)^{-2}|dz|,\label{haarmeasures2}\\
\left.dv(U)\right|_{U(1)^2}&=& -\bar{u}_1du_1 \bar{u}_2du_2.\nn \eea
We have used the Iwasawa decomposition of an element $g$ given in
(\ref{Iwasawa},\ref{Iwasawa2}) and denoted by $|dz|$ and $|dZ|$
the Lebesgue measures on  $\mathbb C$ and $\mathbb C^4$,
respectively (see Appendix \ref{appendixortho} for more explicit expressions of this measure).
The group $U(4)$ is represented in $L^2(U(4),d\mu)$ as (left-action) $[\mathcal{U}(g')\psi](g)=\psi(g'^{-1}g)$. This representation is
reducible and we shall restrict it to an irreducible subspace. As we want to restrict ourselves
to the quotient $U(4)/U(2)^2$, we chose as fiducial  (ground state, lowest weight) vector
$\psi_0^\lambda(g)=\det(D)^\lambda$ for $g$ given in \eqref{Iwasawa} and $\lambda$ an integer number that will eventually
label the corresponding irreducible representation. In fact,
$\psi_0^\lambda(g)$ is invariant (up to a phase) under $U(2)^2\subset U(4)$ since, for $g'=\left(\ba{cc} U_1& 0
\\ 0 &U_2\ea\right)\in U(2)^2$, we have
\be
\psi_0^\lambda(g'^{-1}g)=\det(U_2^\dag D)^\lambda=\det(U_2^\dag)^\lambda\psi_0^\lambda(g).
\ee
Under  a general element $g'=\left(\ba{cc} A'& B'
\\ C' & D'\ea\right)\in U(4)$, the vector $\psi_0^\lambda$ transforms as
\be
\psi_{g'}^\lambda(g)\equiv \psi_0^\lambda(g'^{-1}g)=\det(B'^\dag B+D'^\dag D)^\lambda=
\det(B'^\dag Z+D'^\dag)^\lambda\psi_0^\lambda(g)
,\label{cs}
\ee
where we have used the relations \eqref{zzdag} to write $Z=BD^{-1}$. The set of functions in the orbit of $\psi^\lambda_0$ under $U(4)$
\be
\mathcal{S}_\lambda=\{\psi_{g}^\lambda\equiv\mathcal{U}(g)\psi^\lambda_0,\; g\in U(4)\}
\ee
defines a system of CS.  Note that $\psi_{g}^\lambda$ and $\psi_{g'}^\lambda$ are equivalent (up to a phase) if
$g'g^\dag\in U(2)^2\subset U(4)$. We shall prove that this coherent state system  fulfills the resolution of the identity
\be
1=c_\lambda\int_{\mathbb G_2} \left.d\mu(g)\right|_{\mathbb G_2} |\psi^\lambda_g\rangle \langle \psi^\lambda_g|,\label{resol}
\ee
with a suitable normalization constant $c_\lambda$. Before, let us obtain some auxiliary results.
Note that,  introducing  $Z'^\dag=D'^{\dag-1}B'^\dag$ as in \eqref{zzdag},  the state \eqref{cs} can be written as
\be
\psi_{g'}^\lambda(g)=\det(\sigma_0+Z'^\dag Z)^\lambda\overline{\psi_0^\lambda(g')}\psi_0^\lambda(g).\label{cs2}
\ee
We also realize that $|\psi_0^\lambda(g)|^2=\det(DD^\dag)^\lambda=\det(\sigma_0+Z^\dag Z)^{-\lambda}$. To prove \eqref{resol}, we would like
to have before an expansion of $\det(\sigma_0+Z'^\dag Z)^\lambda$ in terms of orthogonal polynomials. For this purpose, let us
prove an interesting identity that will be useful in the sequel.

{\lem \label{antikernelprop} Let  us denote by
\bea
\cD^{j}_{q_a,q_b}(X)=\sqrt{\frac{(j+q_a)!(j-q_a)!}{(j+q_b)!(j-q_b)!}}
\sum_{k=\max(0,q_a+q_b)}^{\min(j+q_a,j+q_b)}
\binom{j+q_b}{k}\binom{j-q_b}{k-q_a-q_b}\nn\\ \times  x_{11}^k
x_{12}^{j+q_a-k}x_{21}^{j+q_b-k}x_{22}^{k-q_a-q_b},\label{Wignerf}\eea
the usual Wigner's $\cD$-matrices for $SU(2)$ (see e.g. \cite{Louck3}), where $j\in
\mbN/2$ (the spin) runs on all non-negative half-integers and
$q_a,q_b=-j,-j+1,\dots,j-1,j$, and $X$ represents here an arbitrary  $2\times 2$
complex matrix with entries $x_{uv}$. For every $
\lambda  \in \mathbb{N}$  the following identity holds:
 \be
\det(\sigma_0+X)^{\lambda}= \sum^{\lambda}_{m=0}\sum_{j=0;\um}^{(\lambda-m)/2}\frac{2j+1}{\lambda+1}
 \binom{\lambda+1}{2j+m+1}\binom{\lambda+1}{m}
 \det(X)^{m}\sum_{q=-j}^j\cD^{j}_{qq}(X),\label{antikernel}\ee
where the sum on $j$ runs over half-nonnegative integers:  $j=0,\um,1,\frac{3}{2},2,\dots,(\lambda-m)/2$.
}

\ni\textbf{Proof:} We shall proceed by induction on $\lambda$. For $\lambda=1$ we have
\be \det(\sigma_0+X)=1+\tr(X)+\det(X) \label{DTrDet}\nn\ee
with  ${\rm tr}(X)$ and $\det(X)$  homogeneous polynomials of degree
1 and 2 in $x_{uv}$, respectively. Wigner matrices $\cD^j_{qq'}(X)$ are homogeneous polynomials of degree
$2j$ in $x_{uv}$. For the spin-0 singlet representation of $U(2)$ we have $\cD^{0}_{00}(X)=1$ and for the
spin-$1/2$ fundamental representation of $U(2)$ we have
$
 \sum_{q=-1/2}^{1/2}\cD^{1/2}_{qq}(X)=\tr(X),$
and therefore
\bea
\sum^{1}_{m=0}\sum_{j=0;\um}^{(1-m)/2}\frac{2j+1}{2}
 \binom{2}{2j+m+1}\binom{2}{m}
 \det(X)^{m}\sum_{q=-j}^j\cD^{1/2}_{qq}(X)&& \nn\\ =1+\tr(X)+\det(X)=\det(\sigma_0+X).\label{antikernel1}\eea
Thus we proved the identity \eqref{antikernel} for $\lambda=1$. Let us assume that \eqref{antikernel} holds for
some natural $\lambda$.
Inspired by Euler's theorem, we shall define the
following differential operator:
\be D_\lambda \equiv -\lambda+t \frac{\partial}{\partial t},\ee
which will be useful in the sequel. Applying $D_{\lambda+1}$ to $\det(\sigma_0+tX)^{\lambda+1}$ gives
\begin{equation}
D_{\lambda+1}\det(\sigma_0+tX)^{\lambda+1}=-(\lambda+1)\det(\sigma_0+tX)^{\lambda}(1-\det(tX)),\label{derker}
\end{equation}
where we have used that ${\rm tr}(tX)$ and $\det(tX)$  homogeneous polynomials of degree
1 and 2 in the parameter $t$. Assuming now that \eqref{antikernel} holds for some natural 
$\lambda>1$ and inserting it in the r.h.s. of \eqref{derker}, after some algebraic 
manipulations we arrive to
\begin{eqnarray}
&\sum^{\lambda}_{m=0}\sum_{j=0;\um}^{(\lambda-m)/2}{(2j+1)}
 \binom{\lambda+1}{2j+m+1}\binom{\lambda+1}{m}
 (\det(tX)-1)\det(tX)^{m}\sum_{q=-j}^j\cD^{j}_{qq}(tX)&\label{derker2}\\
=&\sum^{\lambda+1}_{m=0}\sum_{j=0;\um}^{(\lambda+1-m)/2}(2j+2m-(\lambda+1))\frac{(2j+1)}{\lambda+2}
 \binom{\lambda+2}{2j+m+1}\binom{\lambda+2}{m}
 \det(tX)^{m}\sum_{q=-j}^j\cD^{j}_{qq}(tX).& \nn
\end{eqnarray}
Taking int account that $\det(tX)^{m}\sum_{q=-j}^j\cD^{j}_{qq}(tX)=t^{2m+2j}
\det(X)^{m}\sum_{q=-j}^j\cD^{j}_{qq}(X)$ (that is, a homogeneous polynomial of degree $2m+2j$ in the 
$X$ entries), we recognize $(2j+2m-(\lambda+1))$ in the r.h.s. of \eqref{derker2} as the eigenvalue of 
$D_{\lambda+1}$. Thus we proved that 
\begin{eqnarray}
&D_{\lambda+1}\det(\sigma_0+tX)^{\lambda+1}&\\ =&D_{\lambda+1}\sum^{\lambda+1}_{m=0}
\sum_{j=0;\um}^{(\lambda+1-m)/2}\frac{(2j+1)}{\lambda+2}
 \binom{\lambda+2}{2j+m+1}\binom{\lambda+2}{m}
 \det(tX)^{m}\sum_{q=-j}^j\cD^{j}_{qq}(tX),& \nn
\end{eqnarray}
which coincides with the result of applying $D_{\lambda+1}$ to both sides of \eqref{antikernel} with $\lambda$ 
replaced by $\lambda+1$. The fact that
$D_\lambda (f(t)+k)=D_\lambda f(t)-\lambda k$, for any constant $k$, eliminates any arbitrarity in $f(t)$. 
Therefore, for $t=1$, we conclude that the equality \eqref{antikernel} is also true for $\lambda+1$, 
thus achieving the proof
by induction.
 $\blacksquare$

Now we are in condition to prove the following interesting result

{\thm\label{GMSMT}  The  set of homogeneous polynomials
\be
\varphi_{q_a,q_b}^{j,m}(Z)=\sqrt{\frac{2j+1}{\lambda+1}\binom{\lambda+1}{2j+m+1}\binom{\lambda+1}{m}}
\det(Z)^{m}\cD^{j}_{q_a,q_b}(Z),\; \begin{matrix}
2j+m\leq\lambda, \\ q_a,q_b=-j,\dots,j, \end{matrix}\label{basisfunc}\ee
 of degree $2j+2m$ verifies the following
closure relation (the reproducing Bergman kernel):
\be\sum^{\lambda}_{m=0}\sum_{j=0;\um}^{(\lambda-m)/2}\sum^{j}_{q_a,q_b=-j}
\overline{\varphi_{q_a,q_b}^{j,m}({Z'})}\varphi_{q_a,q_b}^{j,m}(Z)={\det(\sigma_0+Z'^\dag
Z)^\lambda}\label{closure}\ee
and constitutes an orthonormal basis of the 
\begin{equation}
d_\lambda=(\lambda+1)(\lambda+2)^2(\lambda+3)/12\label{dimensionl}
\end{equation}
dimensional Hilbert space ${\cal H}_\lambda(\mathbb
G_2)=L^2_h(\mathbb G_2,d\mu_\lambda)$ of analytic square-integrable holomorphic functions on $\mathbb G_2$ with measure
\be
d\mu_\lambda(Z,Z^\dag)\equiv c_\lambda|\psi_0^\lambda(g)|^2\left.d\mu(g)\right|_{\mathbb G_2}=c_\lambda\det(\sigma_0+Z^\dag Z)^{-4-\lambda}|dZ|,
\label{projintmeasure}
\ee
where $c_\lambda= {12d_\lambda}/{\pi^4}$ is a normalization constant.}

\ni\textbf{Proof:} Replacing $X=Z'^\dag Z$ in  (\ref{antikernel}) we have
\bea \sum^{\lambda}_{m=0}\sum_{j=0;\um}^{(\lambda-m)/2}\frac{2j+1}{\lambda+1}
 \binom{\lambda+1}{2j+m+1}\binom{\lambda+1}{m}
 \det(Z'^\dag Z)^{m}\sum^{j}_{q=-j}\cD^{j}_{qq}(Z'^\dag Z)\nn\\ =
 {\det(\sigma_0+Z'^\dag Z)^{\lambda }}\label{lemma2}\,.\eea
Using determinant and Wigner's $\cD$-matrix properties \cite{Louck3}
\be
 \det(Z'^\dag Z)^{m}\sum^{j}_{q=-j}\cD^{j}_{qq}(Z'^\dag Z)=
 \det(Z'^\dag)^{m}\det(Z)^{m}\sum^{j}_{q_b,q_a=-j}\overline{\cD^{j}_{q_aq_b}(Z')}\cD^{j}_{q_aq_b}(Z)\nn\ee
and the definition of the functions (\ref{basisfunc}), we see that (\ref{lemma2})
reproduces (\ref{closure}). On the other hand, the number of linearly independent polynomials
$\prod_{i,j=1}^2 z_{ij}^{n_{ij}}$ of fixed degree of homogeneity
$n=\sum_{i,j=1}^2n_{ij}$ is $(n+1)(n+2)(n+3)/6=\binom{n+3}{3}$ (the number of ways of distributing 
$n$ quanta among four levels), which coincides
with the number of linearly independent polynomials
(\ref{basisfunc}) with degree of homogeneity $n=2m+2j$ for $n\leq\lambda$. 
For $\lambda<n=2j+2m\leq 2\lambda$, the degeneracy is $\binom{2\lambda-n+3}{3}$ 
(the number of ways of distributing 
$2\lambda-n$ quanta among four levels). The total number of linearly independent polynomials is
\begin{equation}
 \sum_{n=0}^\lambda \binom{n+3}{3}+ \sum_{n=\lambda+1}^{2\lambda} \binom{2\lambda-n+3}{3}=
 (\lambda+1)(\lambda+2)^2(\lambda+3)/12,
\end{equation}
which coincides with the dimension \eqref{dimensionl}. 
This proves that the set of polynomials (\ref{basisfunc}) is a basis for
analytic functions $\phi\in {\cal H}_\lambda(\mathbb G_2)$. Moreover, this basis turns
out to be orthonormal under the projected integration measure (\ref{projintmeasure}). We address the interested reader
to the Appendix \ref{appendixortho} for details.$\blacksquare$

Let us introduce bracket notation and put
\be
 \la {}{}_{q_a,q_b}^{j,m}|Z\ra \equiv \varphi_{q_a,q_b}^{j,m}(Z)\det(\sigma_0+Z^\dag Z)^{-\lambda/2}.
 \label{Ruhlket}\ee
(We remove the label $\lambda$ from the definition of $|{}{}_{q_a,q_b}^{j,m}\rangle$ for the sake of brevity). This makes  
${\cal H}_\lambda(\mathbb G_2)$ a \emph{reproducing kernel Hilbert space}, that is, a Hilbert space of functions 
$\varphi$ in which pointwise evaluation $\varphi(Z)$ is a continuous linear functional. 
The resolution of the identity for an orthonormal basis in ${\cal H}_\lambda(\mathbb G_2)$ then adopts the form
\be 1=\sum^{\lambda}_{m=0}\sum_{j=0;\um}^{(\lambda-m)/2}\sum^{j}_{q_a,q_b=-j}
|{}{}_{q_a,q_b}^{j,m}\ra \la{}{}_{q_a,q_b}^{j,m}|,
\ee
and the formal ket $|Z\ra$ is
\be
|Z\ra=\det(\sigma_0+Z^\dag Z)^{-\lambda/2}\sum^{\lambda}_{m=0}\sum_{j=0;\um}^{(\lambda-m)/2}\sum^{j}_{q_a,q_b=-j}\varphi_{q_a,q_b}^{j,m}(Z)
|{}{}_{q_a,q_b}^{j,m}\ra.\label{u4cs}
\ee
Actually, we can identify $|Z\ra$ with the coherent state $|\psi^\lambda_g\ra$ up to a phase. From the coherent state overlap
\be
\la Z'|Z\ra=\frac{\det(\sigma_0+Z'^\dag Z)^\lambda}{\det(\sigma_0+Z'^\dag Z')^{\lambda/2}\det(\sigma_0+Z^\dag Z)^{\lambda/2}}\label{u4csov}
\ee
we see that $|Z\ra$ is normalized. Moreover, using the orthogonality properties of the homogeneous polynomials
$\varphi_{q_a,q_b}^{j,m}(Z)$, it is direct to prove the announced resolution of unity \eqref{resol}, now written as:
\be
1=c_\lambda\int_{\mathbb G_2} |Z\ra\la Z|\left.d\mu(g)\right|_{\mathbb G_2}.\label{u4csclos}
\ee
It is interesting to compare the $U(4)/U(2)^2$ CS  \eqref{u4cs} with the well known $U(2)/U(1)^2$  or spin-$s$ CS
\be
|z\rangle=(1+|z|^2)^{-s}\sum_{q=-s}^s\varphi_q(z)|s,q\rangle,\;\; \varphi_q(z)=\binom{2s}{s+q}^{1/2}z^{s+q},\label{su2cs}
\ee
with $z\in\mathbb C$ (the stereographic projection of the sphere $\mathbb S^2=U(2)/U(1)^2$ onto the complex plane), for which the coherent state overlap
and the resolution of the identity acquire the form
\be
\langle z'|z\rangle=\frac{(1+\bar{z'}z)^{2s}}{(1+|z'|^2)^s(1+|z|^2)^s},\;
1=\frac{2s+1}{\pi}\int_{{\mathbb C}}|z\rangle\langle z|\frac{d^2 z}{(1+|z|^2)^2}.\label{overresosu2}
\ee
We perceive a similar structure between $U(4)/U(2)^2$ and $U(2)/U(1)^2$ CS, although the case  $U(4)/U(2)^2$ is more
involved and can be regarded as a  generalized (matrix $Z$) version of the standard (scalar $z$) case.

We finish this section with an explicit form of the unirep of $U(4)$ on $\mathcal{H}_\lambda(\mathbb G_2)$ in the form of a Corollary.

{\cor\label{cor3} For any holomorphic function $\phi\in\mathcal{H}_\lambda(\mathbb G_2)$ and any $g'\in U(4)$, the following
action
\be [{\cal U}^\lambda_{g'}\phi](Z)\equiv \det(D'^\dag+B'^\dag
Z)^{\lambda}\phi(Z'), \; Z'=(A'^\dag Z-C'^\dag)(D'^\dag-B'^\dag Z)^{-1} \label{reprerest}\ee
defines a square-integrable unitary irreducible representation of $U(4)$ on $\mathcal{H}_\lambda(\mathbb G_2)$.
}\\
Note that if we define $\psi(g)\equiv\psi^\lambda_0(g)\phi(Z), Z=Z(g)$, then
\be
[{\cal U}^\lambda_{g'}\phi](Z)=(\psi^\lambda_0(g))^{-1}[\mathcal{U}(g')\psi](g).
\ee
The unitarity of $\mathcal{U}$ in $L^2(U(4),d\mu)$ directly implies the unitarity of ${\cal U}^\lambda$
in $\mathcal{H}(\mathbb G_2)$. Irreducibility follows from the fact that, for example, for $\phi(Z)=1$, the
transformed function
\be
[{\cal U}^\lambda_{g'}\phi](Z)\equiv \det(D'^\dag+B'^\dag
Z)^{\lambda}=\sum^{\lambda}_{m=0}\sum_{j=0;\um}^{(\lambda-m)/2}\sum^{j}_{q_a,q_b=-j}
c_{q_a,q_b}^{j,m}(g')\varphi_{q_a,q_b}^{j,m}(Z)
\ee
is expanded in terms of all basis functions $\varphi_{q_a,q_b}^{j,m}(Z)$ with non-zero coefficients
$c_{q_a,q_b}^{j,m}(g')=\det(D'^\dag)^\lambda \overline{\varphi_{q_a,q_b}^{j,m}({B'D'^{-1}})}$, as follows
from \eqref{closure}.

Our irrep turns out to correspond to the one denoted by the Young Tableau of shape 
$[\lambda,\lambda]$ with two rows of $\lambda$ boxes each (we use the ``English notation'').
This irrep arises in the Clebsch-Gordan decomposition of a tensor product of $N=2\lambda$ four-dimensional (fundamental, elementary) representations of
$U(4)$. The dimension of the tableau  $[\lambda,\lambda]$ can be obtained from the so called ``hook-length'' formula
(which is a special case of the Weyl's character formula, see e.g. \cite{Georgi}) and turns out to coincide with the dimension $d_\lambda$ of  
${\cal H}_\lambda(\mathbb G_2)$ in \eqref{dimensionl}. For example, for $\lambda=1$ ($N=2$ ``particles or quanta'') we have $[1]\otimes [1]=[2] \oplus [1,1]$ or

\be  \begin{Young}  \cr \end{Young}  \otimes \begin{Young}  \cr \end{Young}  =
  \begin{Young} & \cr \end{Young}\oplus \begin{Young} \cr \cr \end{Young}\,\Rightarrow\;\;
 4\times 4=10+6
\ee
so that $[1,1]$ has dimension $d_1=6$.  For $\lambda=2$ ($N=4$ ``particles or quanta'') we have

\be  \begin{Young}  \cr \end{Young}  \otimes \begin{Young}  \cr \end{Young}
\otimes \begin{Young}  \cr \end{Young} \otimes \begin{Young}  \cr \end{Young}=
  \begin{Young} & & & \cr \end{Young}\oplus  \begin{Young} & & \cr \cr \end{Young}\oplus
   \begin{Young} &  \cr \cr \cr \end{Young}\oplus \begin{Young} \cr\cr\cr  \cr \end{Young}\oplus
   \begin{Young} & \cr & \cr \end{Young}
\ee
and the dimension of $[2,2]$ (the last young tableau) is precisely  $d_2=20$. After discussing  an oscillator realization 
of the previous construction later in Section \ref{oscisec}, we will provide in Section \ref{comments} a ``composite fermion'' picture (a term imported from the
quantum Hall effect jargon \cite{Jainbook}) to physically interpret  the $[\lambda,\lambda]$ configurations as two fermions bound to 
$\lambda$ flux quanta each. Before, let us state some interesting results concerning the basic operators and their matrix elements.

\section{Infinitesimal generators and matrix elements}\label{sec3} 

Let us denote by $T_{\mu\nu}$ and $M_{\mu\nu}$ the infinitesimal (differential) generators of the 
finite action \eqref{reprerest} fulfilling the same commutation relations as the 
matrix generators $\tau_{\mu\nu}$ and $\mathfrak{m}_{\mu\nu}$ in \eqref{su4commutators}. 
Writting $Z=z^\mu\sigma_\mu, z^\mu\in\mathbb C$, 
$z^2=z_\mu z^\mu$, $\partial_\mu=\partial/\partial z^\mu$ and 
$\check\partial_\mu=\partial/\partial z_\mu=\partial^\mu$, 
these generators have the following expression:
\bea
{M}_{\mu\nu}=z_\mu \partial_\nu-z_\nu \partial_\mu, &
T_{30}=2(z^\mu\partial_\mu-\lambda),\nn\\
T_{-\mu}=\check\partial_\mu, &
T_{+\mu}= z^2\check T_{-\mu}-z_\mu T_{30},
\eea
where we are using the notation $T_{\pm\mu}=(T_{1\mu}\pm iT_{2\mu})/2$ and $\check T_{\pm\mu}= T_{\pm}^\mu=
\eta^{\mu\nu}T_{\pm\nu}$, as
in \eqref{taupm} and \eqref{Mmunu}. For example, from the general expression
\eqref{reprerest}, we can compute the infinitesimal action of $g'=e^{-it \tau_{30}}$
($B'=0=C'$ and $A'=e^{-it}\sigma_0=D'^\dag$) on wave
functions as $[{\cal U}^\lambda_{g'}\phi](Z)=e^{-2i\lambda t}\phi(e^{2it}Z)=\phi(Z)+it
T_{30}\phi(Z)+O(t^2)$. The other generators are calculated in a similar way. Let us compute
their action on the orthonormal basis functions \eqref{basisfunc}. Firstly we see that the homogeneous polynomials in 
\eqref{basisfunc} are eigenfunctions of the (pseudospin third component) operator  $T_{30}=T_3^0$ since
\be
T_{3}^0\varphi_{q_a,q_b}^{j,m}=2(2j+2m-\lambda)\varphi_{q_a,q_b}^{j,m},\label{interimbalance}
\ee
where the eigenvalue $2(2j+2m-\lambda)$ could be related to an ``imbalance'' or particle 
difference between layers $a$ and $b$ (see next Section). Similarly, we can compute the action of
the lowering interlayer ladder operators ($T_\pm^\mu=\eta^{\mu\nu}T_{\pm\nu}$)
\bea
T_{-}^0\varphi_{q_a,q_b}^{j,m}&=&
C_{q_a,q_b}^{j,m+2j+1}\varphi_{q_a-\um,q_b-\um}^{j-\um,m}+
C_{-q_a+\um,-q_b+\um}^{j+\um,m}\varphi_{q_a-\um,q_b-\um}^{j+\um,m-1}+\nn\\ &&
C_{-q_a,-q_b}^{j,m+2j+1}\varphi_{q_a+\um,q_b+\um}^{j-\um,m}+
C_{q_a+\um,q_b+\um}^{j+\um,m}\varphi_{q_a+\um,q_b+\um}^{j+\um,m-1}\,,\nn\\
T_{-}^1\varphi_{q_a,q_b}^{j,m}&=&
C_{-q_a,q_b}^{j,m+2j+1}\varphi_{q_a+\um,q_b-\um}^{j-\um,m}-
C_{q_a+\um,-q_b+\um}^{j+\um,m}\varphi_{q_a+\um,q_b-\um}^{j+\um,m-1}+\nn\\ &&
C_{q_a,-q_b}^{j,m+2j+1}\varphi_{q_a+\um,q_b-\um}^{j-\um,m}-
C_{-q_a+\um,q_b+\um}^{j+\um,m}\varphi_{q_a-\um,q_b+\um}^{j+\um,m-1}\,,\nn\\
T_{-}^2\varphi_{q_a,q_b}^{j,m}&=&
iC_{-q_a,q_b}^{j,m+2j+1}\varphi_{q_a+\um,q_b-\um}^{j-\um,m}-
iC_{q_a+\um,-q_b+\um}^{j+\um,m}\varphi_{q_a+\um,q_b-\um}^{j+\um,m-1}-\nn\\ &&
iC_{q_a,-q_b}^{j,m+2j+1}\varphi_{q_a+\um,q_b-\um}^{j-\um,m}+
iC_{-q_a+\um,q_b+\um}^{j+\um,m}\varphi_{q_a-\um,q_b+\um}^{j+\um,m-1}\,,\nn\\
T_{-}^3\varphi_{q_a,q_b}^{j,m}&=&
C_{q_a,q_b}^{j,m+2j+1}\varphi_{q_a-\um,q_b-\um}^{j-\um,m}+
C_{-q_a+\um,-q_b+\um}^{j+\um,m}\varphi_{q_a-\um,q_b-\um}^{j+\um,m-1}-\nn\\ &&
C_{-q_a,-q_b}^{j,m+2j+1}\varphi_{q_a+\um,q_b+\um}^{j-\um,m}-
C_{q_a+\um,q_b+\um}^{j+\um,m}\varphi_{q_a+\um,q_b+\um}^{j+\um,m-1}\,,\label{lowering}
\eea
and the raising interlayer ladder operators
\bea
T_{+}^0\varphi_{q_a,q_b}^{j,m}&=&
C_{q_a,q_b}^{j,m+1}\varphi_{q_a-\um,q_b-\um}^{j-\um,m+1}+
C_{-q_a,-q_b}^{j,m+1}\varphi_{q_a+\um,q_b+\um}^{j-\um,m+1}+\nn\\ &&
C_{-q_a+\um,-q_b+\um}^{j+\um,m+2j+1}\varphi_{q_a-\um,q_b-\um}^{j+\um,m}+
C_{q_a+\um,q_b+\um}^{j+\um,m+2j+1}\varphi_{q_a+\um,q_b+\um}^{j+\um,m}\,,\nn\\
T_{+}^1\varphi_{q_a,q_b}^{j,m}&=&
C_{-q_a+\um,q_b+\um}^{j+\um,m+2j+1}\varphi_{q_a-\um,q_b+\um}^{j+\um,m}+
C_{q_a+\um,-q_b+\um}^{j+\um,m+2j+1}\varphi_{q_a+\um,q_b-\um}^{j+\um,m}-\nn\\ &&
C_{q_a,-q_b}^{j,m+1}\varphi_{q_a-\um,q_b+\um}^{j-\um,m+1}-
C_{-q_a,q_b}^{j,m+1}\varphi_{q_a+\um,q_b-\um}^{j-\um,m+1}\,,
\nn\\
T_{+}^2\varphi_{q_a,q_b}^{j,m}&=&
-iC_{-q_a+\um,q_b+\um}^{j+\um,m+2j+1}\varphi_{q_a-\um,q_b+\um}^{j+\um,m}+
iC_{q_a+\um,-q_b+\um}^{j+\um,m+2j+1}\varphi_{q_a+\um,q_b-\um}^{j+\um,m}+\nn\\ &&
iC_{q_a,-q_b}^{j,m+1}\varphi_{q_a-\um,q_b+\um}^{j-\um,m+1}-
iC_{-q_a,q_b}^{j,m+1}\varphi_{q_a+\um,q_b-\um}^{j-\um,m+1}\,,
\nn\\
T_{+}^3\varphi_{q_a,q_b}^{j,m}&=&
C_{q_a+\um,q_b+\um}^{j+\um,m+2j+1}\varphi_{q_a+\um,q_b+\um}^{j+\um,m}-
C_{-q_a+\um,-q_b+\um}^{j+\um,m+2j+1}\varphi_{q_a-\um,q_b-\um}^{j+\um,m}+\nn\\ &&
C_{-q_a,-q_b}^{j,m+1}\varphi_{q_a+\um,q_b+\um}^{j-\um,m+1}-
C_{q_a,q_b}^{j,m+1}\varphi_{q_a-\um,q_b-\um}^{j-\um,m+1}\,,\label{rising}
\eea
with
\be
C_{q_a,q_b}^{j,m}=\frac{\sqrt{(j+q_a)(j+q_b)m(\lambda-(m-2))}}{\sqrt{2j(2j+1)}}.
\ee
The differential representation of the top and bottom layer angular momentum in \eqref{angmom} is 
${S}_{aj}=\frac{1}{4}(\check{T}_{0j}+\check{T}_{3j})=\um(M_{0j}-i\epsilon_{jkl}M_{kl})$
and ${S}_{bj}=\frac{1}{4}({T}_{0j}-{T}_{3j})=\um(M_{0j}+i\epsilon_{jkl}M_{kl})$. The action of the spin third component is 
\be
{S}_{\ell 3}\,\varphi_{q_a,q_b}^{j,m}=q_\ell\, \varphi_{q_a,q_b}^{j,m},\; \ell=a, b \label{jzeta}
\ee
and the action of the ladder spin operators is 
\bea
{S}_{\ell\pm}\,\varphi_{q_a,q_b}^{j,m}=\sqrt{(j\mp q_\ell)(j\pm q_\ell+1)}\,
\varphi_{q_a\pm\delta_{\ell,a},q_b\pm\delta_{\ell,b}}^{j,m},\; \ell=a,b \label{jpm}
\eea
where ${S}_{a\pm}={S}_{a 1}\mp i{S}_{a 2}$ and  ${S}_{b\pm}=
{S}_{b 1}\pm i{S}_{b 2}$. Note that ${S}_{a\pm}$ and ${S}_{b\pm}$
have conjugated definitions ($\pm\leftrightarrow\mp$).
This fact is related to the transformation property of wave functions in \eqref{reprerest} which,
for pure rotations ($C'=0=B', \,
A'=V_a, D'=V_b; V_\ell\in SU(2), \ell=a,b$) gives $[{\cal U}^\lambda_{g'}\phi](Z)=\phi(V_a^\dag ZV_b)$, so that rotations 
$V_a$ on the layer $a$ are 
represented by the inverse $V_a^\dag$. This fact resembles 
the difference between space-fixed and body-fixed rigid-rotor angular momentum operators, as commented after
equation \eqref{angmom}.

For completeness, we also give the action of $U(2)^2$-invariant (i.e., commuting with $M_{\mu\nu}$) 
quadratic operators: $M^2= M_{\mu\nu} M^{\mu\nu}$,
$T_{\pm}T_{\pm}=T_{\pm\mu}T_{\pm}^\mu$,  $\check T_{+} T_{-}=
\check T_{+\mu} T_-^\mu$ and $ \check T_{-} T_{+}=
\check T_{-\mu}  T_{+}^{\mu}$,
which results in
\bea
 M^2 \varphi_{q_a,q_b}^{j,m}&=&-8j(j+1)\varphi_{q_a,q_b}^{j,m},\nn\\
T_{-}T_-\varphi_{q_a,q_b}^{j,m}&=& 4\sqrt{m(2j+m+1)(\lambda-m+2)(\lambda-2j-m+1)}
\varphi_{q_a,q_b}^{j,m-1},\nn\\
T_{+}T_+\varphi_{q_a,q_b}^{j,m}&=& 4\sqrt{(m+1)(2j+m+2)(\lambda-m+1)(\lambda-2j-m)}
\varphi_{q_a,q_b}^{j,m+1},\nn\\
\check T_{+} T_{-}\varphi_{q_a,q_b}^{j,m}&=& -4(2j^2+m(m-\lambda-2)+j(2m-\lambda-1))
\varphi_{q_a,q_b}^{j,m},\nn\\
\check T_{-} T_{+}\varphi_{q_a,q_b}^{j,m}&=& -4(2j^2+(m+2)(m-\lambda)+j(2m-\lambda+3))
\varphi_{q_a,q_b}^{j,m}.\label{ppkk}
\eea
With these ingredients, the value of the quadratic Casimir operator \eqref{Casimir}  (written 
in terms of $T_{\mu\nu}$) in the
Hilbert space $\mathcal{H}_\lambda(\mathbb G_2)$ is easily computed and gives:
\be
\mathcal C_2\varphi_{q_a,q_b}^{j,m}=\lambda(\lambda+4)\varphi_{q_a,q_b}^{j,m},\;\;\forall j,m,q_a,q_b.\label{casieigen}
\ee

\section{Oscillator realization}\label{oscisec}

It is well known the oscillator (Schwinger) realization of the $SU(2)$ angular momentum operators $\mathcal{S}_z, \mathcal{S}_\pm$ in terms
of two bosonic modes $a$ and $b$ as
\be
\mathcal{S}_z=\um(a^\dag a-b^\dag b),\; \mathcal{S}_+=a^\dag b,\; \mathcal{S}_-=b^\dag a,\label{schwingersu2}
\ee
and the expression of spin-$s$ basis states $|s,q\ra, \, q=-s,\dots,s$, in terms of Fock states
($|0\ra$ denotes the Fock vacuum)
\be
|n_a\ra\otimes |n_b\ra=\frac{(a^\dag)^{n_a}(b^\dag)^{n_b}}{\sqrt{n_a!n_b!}}|0\ra\label{Focku2}
\ee
as
\be
|s,q\ra=\frac{(a^\dag)^{s+q}(b^\dag)^{s-q}}{\sqrt{(s+q)!(s-q)!}}|0\ra=
\frac{\varphi_q(a^\dag)}{\sqrt{\frac{(2s)!}{(s+q)!}}}\frac{\varphi_{-q}(b^\dag)}{\sqrt{\frac{(2s)!}{(s-q)!}}}|0\ra=|s+q\ra_a\otimes |s-q\ra_b,\label{basisinfocksu2}
\ee
where we have used the monomials 
$\varphi_q$ in \eqref{su2cs} as operator functions, since this notation will be generalized in a natural way 
later in eq. \eqref{basisinfock2} for a Fock representation of the basis functions $|{}{}_{q_a,q_b}^{j,m}\ra$ of 
${\cal H}_\lambda(\mathbb G_2)$. Note that the total number of quanta is fixed to $n_a+n_b=(s+q)+(s-q)=2s$. 
The
lowest weight state $|s,-s\ra=\frac{(b^\dag)^{2s}}{\sqrt{(2s)!}}|0\ra$ is often regarded as a {\it
boson condensate} and the rest of states $|s,q\ra$ as excitations above this condensate. The
$SU(2)$ spin-$s$ coherent state \eqref{su2cs} can also be written as
\be
|z\ra=\frac{1}{\sqrt{(2s)!}}\left(\frac{b^\dag+za^\dag}{\sqrt{1+|z|^2}}\right)^{2s}|0\ra=\frac{e^{z \mathcal{S}_+}}{(1+|z|^2)^{s}}|s,-s\rangle
.\label{su2csboson}
\ee
The natural generalization to $U(4)$ requires four bosonic modes $a, b, c$ and $d$, for which the basis
states
\be
|n_a\ra\otimes |n_b\ra\otimes |n_c\ra\otimes |n_d\ra=
\frac{(a^\dag)^{n_a}(b^\dag)^{n_b}(c^\dag)^{n_c}(d^\dag)^{n_d}}{\sqrt{n_a!n_b!n_c!n_d!}}|0\ra, \label{symmetricbasis}
\ee
with $n_a+n_b+n_c+n_d=N $ the total (fixed, linear Casimir) number of ``particles or quanta'', 
all belong to the totally symmetric irreducible representation of $U(4)$. This representation is related
to the quotient $\mathbb CP^3=U(4)/U(3)\times U(1)$ (the complex projective space) whose points
$z_a, z_b, z_c\in \mathbb C$ (in a certain patch) label the CS
\be
|z_a,z_b,z_c\ra=\frac{1}{\sqrt{N!}}\left(
\frac{d^\dag+z_cc^\dag+z_bb^\dag+z_aa^\dag}{\sqrt{1+|z_a|^2+|z_b|^2+|z_c|^2}}\right)^{N}|0\ra.\label{bosoncsu4rpoj}
\ee
These CS
also verify a resolution of the identity similar to the one in \eqref{overresosu2} but replacing the
$\mathbb CP^1$ integration measure by the corresponding $\mathbb CP^3$ integration measure. Fields taking values in the target manifold
$\mathbb CP^3$ describe Goldstone bosons, $SU(4)$-skyrmions and small fluctuations around the ground
state in the bilayer quantum Hall system
at filling factor $\nu=1$ \cite{EzawaBook}.

However,
these are not the CS \eqref{u4cs} we are dealing with in this article. Actually, the CS \eqref{u4cs}
will be related to the filling factor $\nu=2$ in the BLQH system. The question is:
is there a boson realization like \eqref{bosoncsu4rpoj} but for the CS
\eqref{u4cs} labeled by points $Z$ in the
complex Grassmannian $\mathbb G_2$?. The answer is positive and it will be given later in Proposition
\ref{propcsboson}.

The most popular oscillator realization  of the Lie algebra $u(n)$ is that in terms of bilinear products of $n$ creation and annihilation
operators (Schwinger representation) leading to the totally symmetric representation  (for example, the Bose-Einstein-Fock basis
\eqref{symmetricbasis} for $n=4$). Although perhaps less known, other realizations of $u(n)$ in terms of more that $n$ bosonic modes have also been used in the literature
\cite{MoshinskyPL,MoshinskyBook,MoshinskyJMP,MoshinskyNPB}, which describe more general representations than the symmetric one. 
Let us provide an oscillator realization for the (non-symmetric)
$U(4)$ representation given in the previous sections.

Note that, defining $\mathcal Z=\begin{pmatrix} a\\ b \end{pmatrix}$ and
$\mathcal Z^\dag=\begin{pmatrix} a^\dag & b^\dag \end{pmatrix}$, the angular momentum operators
\eqref{schwingersu2} can be compactly written as
\be
\mathcal{S}_\mu=\um\tr(\mathcal Z^\dag\sigma_\mu \mathcal Z), \ee
with
$\mathcal{S}_\pm=\mathcal{S}_1\pm i\mathcal{S}_2$, $\mathcal{S}_z=\mathcal{S}_3$ and $\mathcal Z^\dag\mathcal Z=2\mathcal{S}_0=a^\dag a+b^\dag b$ the
total number of quanta, which is fixed to $N=2s$. This construction
can be straightforwardly extended to $u(4)$ by defining now
\be
\mathcal Z=\begin{pmatrix}
            \mathbf a\\ \mathbf b
           \end{pmatrix}=
\begin{pmatrix}
\begin{matrix} a_0 & a_1\\ a_2 & a_3
\end{matrix}
\\ \begin{matrix} b_0 & b_1\\ b_2 & b_3
\end{matrix} \end{pmatrix}.\label{calzeta}\ee
The oscillator realization of the $u(4)$ generators $\tau_{\mu\nu}$ is given by
\be
\mathcal{T}_{\mu\nu}=\tr({\mathcal Z}^\dag\tau_{\mu\nu} \mathcal Z).\label{bosrepre} \ee
Indeed, one can easily verify that $[\mathcal{T}_{\mu\nu},\mathcal{T}_{\mu'\nu'}]=\tr(\mathcal Z^\dag
[\tau_{\mu\nu},\tau_{\mu'\nu'}] \mathcal Z)$, and therefore \eqref{bosrepre} defines a
(unitary) representation of $u(4)$ in the Fock space
\be
|\mathbf n_a\ra \otimes |{\mathbf n}_b\ra=\left|\begin{matrix} n_a^0 & n_a^1\\ n_a^2 & n_a^3
\end{matrix}\right>\otimes \left|\begin{matrix} n_b^0 & n_b^1\\ n_b^2 & n_b^3
\end{matrix}\right>=\prod_{\mu=0}^3
\frac{(a^\dag_\mu)^{n_a^\mu}(b^\dag_\mu)^{n_b^\mu}}{\sqrt{n_a^\mu!n_b^\mu!}}|0\ra.\label{grassmannbasis2}
\ee

Let us look for the expression of the basis states $|{}{}_{q_a,q_b}^{j,m}\ra$ in \eqref{Ruhlket} in terms of the
Fock basis \eqref{grassmannbasis2}. It is clear that some constraints must be imposed to the
occupancy numbers $n_a^\mu$ and $n_b^\mu$ in order to obtain a $d_\lambda$-dimensional Hilbert space.
In particular, we shall see that the constraint
$\mathcal Z^\dag \mathcal Z=\mathbf a^\dag \mathbf a+\mathbf b^\dag \mathbf b=\lambda
I_{2\times 2}$ is fulfilled on the basis states $|{}{}_{q_a,q_b}^{j,m}\ra$, where $I_{2\times 2}$
denotes the $2\times 2$ identity operator.
Firstly we have to fix the total number of quanta $\sum_{\mu=0}^3 n_a^\mu+n_b^\mu=2\lambda$, that is,
the linear Casimir operator $\mathcal{T}_{00}=\sum_{\mu=0}^3 a^{\dag}_\mu a_\mu+b^{\dag}_\mu b_\mu$
is fixed to $2\lambda$. From \eqref{interimbalance}, we also see that the interlayer imbalance operator
$\mathcal{T}_{30}=\sum_{\mu=0}^3 a^{\dag}_\mu a_\mu-b^{\dag}_\mu b_\mu$ provides the relation
$\sum_{\mu=0}^3 (n_a^\mu-n_b^\mu)=2(2j+2m-\lambda)$, so that, when the homogeneity degree $(2j+2m)$ of
$\varphi^{j,m}_{q_a,q_b}$ equals $\lambda$ (half the total number of quanta), the configuration
$|{}{}_{q_a,q_b}^{j,m}\ra$ is balanced (same number of quanta in both layers $a$ and $b$). Therefore,
the lowest-weight (zero homogeneity degree) state  $|\varphi_0\ra\equiv|{}{}_{q_a=0,q_b=0}^{j=0,m=0}\ra$ is made of $2\lambda$ quanta 
in the bottom layer $b$ and can expressed in terms
of Fock states as:
\be
|\varphi_0\ra=\frac{\det(\mathbf b^\dag)^\lambda}{\lambda!\sqrt{\lambda+1}}|0\ra= \left|\begin{matrix} 0 & 0\\ 0 & 0
\end{matrix}\right>_a\otimes \sum_{k=0}^\lambda\frac{ (-1)^k}{\sqrt{\lambda+1}}\left|\begin{matrix}
\lambda-k & k\\ k & \lambda-k
\end{matrix}\right>_b.\label{lowestweight}\ee
Indeed, one can easily check that $|\varphi_0\ra$ fulfills the
constraint $\mathcal Z^\dag \mathcal Z=\lambda
I_{2\times 2}$.

Applying ladder operators (\ref{lowering},\ref{rising},\ref{jpm}) and \eqref{ppkk} to the lowest-weight state 
\eqref{lowestweight} we have been able to obtain the expression of the basis states $|{}{}_{q_a,q_b}^{j,m}\ra$
in terms of Fock states \eqref{grassmannbasis2} step by step.\footnote{We do not present here the (rather 
cumbersome) steps to get this result. We must acknowledge the benefits of  \emph{Mathematica} add-on packages like 
``Quantum Algebra'' to check this and some other 
expressions along this Section. These packages are available at \cite{Quantum}.} In the process we find extra restrictions to the
number $n_a^\mu$ and $n_b^\mu$ of quanta in layers $a$ and $b$ like:
\be
n_a^0+n_a^1+n_a^2+n_a^3=2(j+m),
\ee
which says that the homogeneity degree $2(j+m)$ of $\varphi^{j,m}_{q_a,q_b}$ represents the total number of quanta 
in the top layer $a$. Other restriction is 
\be
n_a^0+n_a^2+n_b^0+n_b^2=\lambda=n_a^1+n_a^3+n_b^1+n_b^3\,,
\ee
which states that the total number of ``even'' ($\mu=0,2$) and ``odd''  ($\mu=1,3$) quanta in both layers must be
balanced. In the ``composite bi-fermion'' picture \eqref{calzetaSky} of the next Section, ``even and odd'' (flux) quanta are 
attached to the ``first and second'' fermions, respectively. Another interesting restriction is
\bea
n_a^0+n_a^1-n_a^2-n_a^3&=&-2q_a\,,\nn\\
n_b^0+n_b^1-n_b^2-n_b^3&=&2q_b\,,
\eea
which says that the ``magnetic quantum numbers'' $q_a$ and $q_b$, measure the imbalance between $\mu=\{0,1\}$ (spin up) and
$\mu=\{2,3\}$ (spin down) type ``flux'' quanta (see next Section for a physical interpretation) inside layers $a$ and $b$, respectively. 
Note the difference of sign in the definition of $q_a$ and $q_b$.

The final expression of the basis states $|{}{}_{q_a,q_b}^{j,m}\ra$
in terms of Fock states \eqref{grassmannbasis2} is
\be
|{}{}_{q_a,q_b}^{j,m}\ra=\frac{1}{\sqrt{2j+1}}\sum_{q=-j}^{j}(-1)^{q_a-q}
|v_{-q,-q_a}^{j,m}\ra_a \otimes
|{v}_{q,q_b}^{j,\lambda-2j-m}\ra_b,\label{basisinfock}
\ee
where
\be
|v_{q,q'}^{j,m}\ra=\sum_{k=\max(0,q+q')}^{j+m+\min(q,q')}G^{j,m}_{q,q'}(k)
\left|\begin{matrix} k & j+m+q'-k\\ j+m+q-k & k-q-q'
\end{matrix}\right>,\label{uves}
\ee
(either for layers $a$ and $b$) with
\bea
G^{j,m}_{q,q'}(k)&=&\sqrt{\frac{2j+1}{(2j+m+1)!m!}\,
\frac{(j+q)!(j-q)!}{(j+q')!(j-q')!}}\nn\\ && \times
\sqrt{(j+m+q-k)!(j+m+q'-k)!(k-q-q')!k!}\nn\\ && \times \sum_{p=0}^m(-1)^p \binom{j+q'}{k-m+p}
\binom{j-q'}{k-m+p-q-q'} \binom{m}{p}\,.
\eea
As the simplest example, let us provide the explicit expression of the basis states $|{}{}_{q_a,q_b}^{j,m}\ra$ for
two quanta ($\lambda=1$):
\bea
|{}{}_{0,0}^{0,0}\ra&=&\frac{1}{\sqrt{2}}\left(
\left|\begin{matrix} 0 & 0\\ 0 & 0
\end{matrix}\right>_a\otimes \left|\begin{matrix} 1 & 0\\ 0 & 1
\end{matrix}\right>_b-
\left|\begin{matrix} 0 & 0\\ 0 & 0
\end{matrix}\right>_a\otimes  \left|\begin{matrix} 0 & 1\\ 1 & 0
\end{matrix}\right>_b
\right),\nn\\
|{}{}_{\um,\um}^{\um,0}\ra&=&\frac{1}{\sqrt{2}}\left(
\left|\begin{matrix} 0 & 0\\ 0 & 1
\end{matrix}\right>_a\otimes \left|\begin{matrix} 1 & 0\\ 0 & 0
\end{matrix}\right>_b-
\left|\begin{matrix} 0 & 0\\ 1 & 0
\end{matrix}\right>_a\otimes  \left|\begin{matrix} 0 & 1\\ 0 & 0
\end{matrix}\right>_b
\right),\nn\\
|{}{}_{-\um,-\um}^{\;\;\um,\;\;0}\ra&=&\frac{1}{\sqrt{2}}\left(
\left|\begin{matrix} 1 & 0\\ 0 & 0
\end{matrix}\right>_a\otimes \left|\begin{matrix} 0 & 0\\ 0 & 1
\end{matrix}\right>_b-
\left|\begin{matrix} 0 & 1\\ 0 & 0
\end{matrix}\right>_a\otimes  \left|\begin{matrix} 0 & 0\\ 1 & 0
\end{matrix}\right>_b
\right),\nn\\
|{}{}_{-\um,\um}^{\;\um,\,0}\ra&=&\frac{1}{\sqrt{2}}\left(
\left|\begin{matrix} 1 & 0\\ 0 & 0
\end{matrix}\right>_a\otimes \left|\begin{matrix} 0 & 1\\ 0 & 0
\end{matrix}\right>_b-
\left|\begin{matrix} 0 & 1\\ 0 & 0
\end{matrix}\right>_a\otimes  \left|\begin{matrix} 1 & 0\\ 0 & 0
\end{matrix}\right>_b
\right),\nn\\
|{}{}_{\um,-\um}^{\,\um,\;0}\ra&=&\frac{1}{\sqrt{2}}\left(\left|\begin{matrix} 0 & 0\\ 0 & 1
\end{matrix}\right>_a\otimes  \left|\begin{matrix} 0 & 0\\ 1 & 0
\end{matrix}\right>_b-
\left|\begin{matrix} 0 & 0\\ 1 & 0
\end{matrix}\right>_a\otimes \left|\begin{matrix} 0 & 0\\ 0 & 1
\end{matrix}\right>_b
\right),\nn\\
|{}{}_{0,0}^{0,1}\ra&=&\frac{1}{\sqrt{2}}\left(\left|\begin{matrix} 1 & 0\\ 0 & 1
\end{matrix}\right>_a\otimes  \left|\begin{matrix} 0 & 0\\ 0 & 0
\end{matrix}\right>_b-
\left|\begin{matrix} 0 & 1\\ 1 & 0
\end{matrix}\right>_a\otimes \left|\begin{matrix} 0 & 0\\ 0 & 0
\end{matrix}\right>_b
\right).
\eea

One can prove that the set of vectors $|v_{q,q'}^{j,m}\ra$ constitutes an orthonormal set for each layer, that is
\be
\la v_{q_a,q}^{j,m}|v_{q_a',q'}^{j',m'}\ra=\delta_{j,j'}\delta_{m,m'}\delta_{q_a,q_a'}\delta_{q,q'}.\label{ortholayer}
\ee

After some algebra, one can realize that the states \eqref{uves} can be obtained as
\be
|v_{q,q'}^{j,m}\ra_a=\sqrt{\frac{(\lambda-2j-m)!(\lambda+1-m)!}{\lambda!(\lambda+1)!}}\;\varphi^{j,m}_{q,q'}(\mathbf{a}^\dag)|0\ra,
\ee
(and an equivalent expression for the layer $b$), where we are treating now the homogeneous
polynomials $\varphi^{j,m}_{q,q'}$ in \eqref{basisfunc} as operator functions, since there is not
ordering problem (all $a_\mu^\dag$ and $b_\mu^\dag$ commute). Therefore, the basis
states \eqref{basisinfock} can be obtained from the Fock vacuum $|0\ra$ as
\be
|{}{}_{q_a,q_b}^{j,m}\ra=\frac{1}{\sqrt{2j+1}}\sum_{q=-j}^{j}(-1)^{q_a-q}
\frac{\varphi^{j,m}_{-q,-q_a}(\mathbf{a}^\dag)}{\sqrt{\frac{\lambda!(\lambda+1)!}{(\lambda-2j-m)!(\lambda+1-m)!}}}
\frac{\varphi^{j,\lambda-2j-m}_{q,q_b}(\mathbf{b}^\dag)}{\sqrt{\frac{\lambda!(\lambda+1)!}{m!(2j+m+1)!}}}
\;|0\ra.
\label{basisinfock2}
\ee
This is the $SU(4)$ version of eq. \eqref{basisinfocksu2} for the spin-$s$ basis states $|s,q\ra$ of $SU(2)$,
with the role of the spin $s$ played now by $\lambda$ and the role of the monomials $\varphi_q(z)$ played now by the 
homogeneous polynomials $\varphi_{q_a,q_b}^{j,m}(Z)$. 

At this point, we are in condition to provide a boson realization
like \eqref{su2csboson} and \eqref{bosoncsu4rpoj} but for the CS
\eqref{u4cs} labeled by points $Z$ in
complex Grassmannian $\mathbb G_2$

{\prop\label{propcsboson} Let us denote by $\check{\mathbf{a}}=\um\eta^{\mu\nu}\tr(\sigma_\mu\mathbf{a})\sigma_\nu$ and
$\check{\mathbf{b}}=\um\eta^{\mu\nu}\tr(\sigma_\mu\mathbf{b})\sigma_\nu$. The
CS $|Z\ra$ in  \eqref{u4cs} can be written as a boson condensate
\be
|Z\ra=\frac{1}{\lambda!\sqrt{\lambda+1}}\left(\frac{\det(\check{\mathbf{b}}^\dag+
Z^t\check{\mathbf{a}}^\dag)}{\sqrt{\det(\sigma_0+Z^\dag Z)}}\right)^\lambda|0\ra.
\label{u4csfock}
\ee
}

\noindent\textbf{Proof:} Using similar steps as in the proof of Lemma \ref{antikernelprop}
and Theorem \ref{GMSMT}, we can also proof that, for any $2\times 2$ matrices $A, B$ and
$C$ with $A$ invertible, the following identity holds
\be
\frac{\det(A+BC)^\lambda}{\lambda!\sqrt{\lambda+1}}=\sum^{\lambda}_{m=0}\sum_{j=0;\um}^{(\lambda-m)/2}\sum^{j}_{q_a,q_b=-j}
V_{q_a,q_b}^{j,m}(A,B)\varphi_{q_a,q_b}^{j,m}(C),
\ee
with
\be
V_{q_a,q_b}^{j,m}(A,B)=\frac{1}{\sqrt{2j+1}}\sum_{q=-j}^j (-1)^{q_b-q}
\frac{\varphi^{j,m}_{q,q_a}(B)}{\sqrt{\frac{\lambda!(\lambda+1)!}{(\lambda-2j-m)!(\lambda+1-m)!}}}
\frac{\varphi^{j,\lambda-2j-m}_{-q,-q_b}(A)}{\sqrt{\frac{\lambda!(\lambda+1)!}{m!(2j+m+1)!}}}.
\ee
Taking into account the following properties
\be
A^{-1}=\frac{\check{A}}{\det(A)},\; \det(\check{A})=\det(A^t), \; \cD^{j}_{q_a,q_b}(\check{X})=
(-1)^{2j+q_a+q_b}\cD^{j}_{-q_a,-q_b}(X^t)
\ee
and identifying $A^t\to \check{\mathbf{b}}^\dag$, $B^t\to \check{\mathbf{b}}^\dag$ and
$C\to Z$, the expression \eqref{u4csfock} reduces to \eqref{u4cs} through the identification
\eqref{basisinfock2}$\blacksquare$

For $Z=0$ we recover the lowest-weight state $|\varphi_0\ra$ in eq. \eqref{lowestweight} since
$\det(\check{\mathbf{b}}^\dag)=\det({\mathbf{b}}^\dag)$.

To finish, let us provide another expression of the
CS $|Z\ra$ in  \eqref{u4cs}, now as an exponential of creation operators.

{\prop\label{propcsboson2} Let us denote by $\mathcal{T}_{+}\equiv
\mathcal{T}_{+}^{\mu}\sigma_\mu=2\check{\mathbf{a}}^\dag\check{\mathbf{b}}$. The
CS $|Z\ra$ in  \eqref{u4cs} and \eqref{u4csfock} can be written as the exponential action on the lowest-weight state
\be
|Z\ra=\frac{e^{\um \tr(Z^t\mathcal{T}_{+})}}{\det(\sigma_0+Z^\dag Z)^{\lambda/2}}|\varphi_0\ra.\label{u4cs2}
\ee
}
Proving \eqref{u4cs2} is equivalent to prove that
\be
e^{\um \tr(Z^t\mathcal{T}_{+})}|\varphi_0\ra=\sum^{\lambda}_{m=0}\sum_{j=0;\um}^{(\lambda-m)/2}
\sum^{j}_{q_a,q_b=-j}\varphi_{q_a,q_b}^{j,m}(Z)
|{}{}_{q_a,q_b}^{j,m}\ra,
\ee
which can be done by induction on the homogeneity degree in $Z$. We shall not give here the (rather cumbersome)
details and only shall point out that the equivalence of the expressions \eqref{u4cs}, \eqref{u4csfock} and \eqref{u4cs2}
for CS on $U(4)/U(2)^2$ is the counterpart of the equivalence of \eqref{su2cs} and \eqref{su2csboson} for
CS on $U(2)/U(1)^2$.

\section{Physical interpretation and some comments\label{comments}}

Let us propose a physical interpretation of the previous abstract mathematical  construction by making use 
of the fractional QH effect notion of \emph{composite fermion} \cite{Jainbook}. 
The  composite-fermion (CF) theory maps the strongly interacting system of 
electrons in a partially filled Landau level to a system of weakly
interacting particles called composite fermions, which are
bound states of an electron and a certain number of flux quanta (quantized
vortices).  The hierarchy of fractional QH states 
is understood by the use of composite fermions. Bilayer composite fermion states 
have also been studied \cite{Jainbook}. Here we shall try to make compatible our construction 
with the composite fermion picture of the BLQH system at filling factor $\nu=2$ and its fractions.

In the BLQH system at filling factor $\nu=2$, there are two electrons in one Landau site. 
Charged excitations are bi-Skyrmions
in the $\nu=2$ BLQH system \cite{Ezawabisky}. 
The $\mathbb G_2$-Skyrmion has the general expression

\be
\mathfrak{Z}=\begin{pmatrix}
           \mathfrak{{z}}_1 & \mathfrak{z}_2
           \end{pmatrix}=
\begin{pmatrix}
\begin{matrix} \mathfrak{z}_1^{a\uparrow} & \mathfrak{z}_2^{a\uparrow}\\ \mathfrak{z}_1^{a\downarrow} & \mathfrak{z}_2^{a\downarrow}
\end{matrix}
\\ \begin{matrix} \mathfrak{z}_1^{b\uparrow} & \mathfrak{z}_2^{b\uparrow}\\ \mathfrak{z}_1^{b\downarrow} & \mathfrak{z}_2^{b\downarrow}
\end{matrix} \end{pmatrix},\label{calzetaSky}\ee
where  $\mathfrak{z}_1$ and 
$\mathfrak{z}_1$ are two $\mathbb CP^3$ fields orthogonal one to another $\mathfrak{z}_1\cdot \mathfrak{z}_2=0$. The 
reader can note the similarity between the bi-Skyrmion \eqref{calzetaSky} and the bosonic matrix \eqref{calzeta}. 
Though there are two fields, $(\mathfrak{z}_1, \mathfrak{z}_2)$, we
cannot distinguish them quantum mechanically since they describe two electrons in the same Landau site. 
Thus, $\mathfrak{Z}$ is not exactly a set of two independent $\mathbb CP^3$ fields. In fact, 
two fields $\mathfrak{Z}$ and $\mathfrak{Z}'$ are indistinguishable when they are related by a local $V\in U(2)$ 
transformation $\mathfrak{Z}'=\mathfrak{Z}V$. The identification $\mathfrak{Z}'\sim \mathfrak{Z}$ leaves 
only four complex field degrees of freedom  $Z=z^\mu\sigma_\mu,\, z^\mu\in\mathbb C$, $\mu=0,1,2,3$. Here we have 
restricted to one Landau site of the Lowest Landau Level. 
Hence the parameter space characterizing the $U(4)$ invariant ground state in the BLHQ system at $\nu=2$ contains 
four complex independent variables.  They are also the four complex Goldstone modes associated with a 
spontaneous breakdown of the  $U(4)$ symmetry.

For fractional filling factors $\nu=\frac{2}{\lambda}$ we can think of the following ``composite bi-fermion'' picture. 
We have two electrons attached to $\lambda$ flux quanta each.
The first electron can occupy any of the four isospin states $|b\uparrow\rangle, |b\downarrow\rangle, 
|a\uparrow\rangle$ and $|a\downarrow\rangle$ in the lowest Landau level. Therefore, there are $\binom{\lambda+3}{3}$ ways of 
distributing $\lambda$ quanta among these four states. Due to the Pauli exclusion principle, there are only three states 
left for the second electron and $\binom{\lambda+2}{2}$ ways of 
distributing $\lambda$ quanta among these three states. However, some of the previous configurations must be identified 
since both electrons are indistinguishable and $\lambda$ pairs of quanta addopt $\binom{\lambda+1}{1}$ 
equivalent configurations. In total, there are 
\begin{equation}
 \frac{\binom{\lambda+3}{3}\binom{\lambda+2}{2}}{\binom{\lambda+1}{1}}=\frac{1}{12}(\lambda+3)(\lambda+2)^2(\lambda+1)
\end{equation}
ways to distribute $2\lambda$ flux quanta among two identical electrons in four states, which turns out to coincide with 
the dimension $d_\lambda$ in \eqref{dimensionl} of the Hilbert space  ${\cal H}_\lambda(\mathbb
G_2)$ of analytic square-integrable holomorphic functions on $\mathbb G_2$ introduced in Theorem \ref{GMSMT}. Using 
Haldane's sphere picture \cite{Haldane} for the fractional QH effect, $\lambda$ is also related to the ``monopole 
strength'' in $\mathbb G_2$. Like Haldane's sphere for monolayer systems, we believe that our construction on 
$\mathbb G_2$ will be very convenient for analytical studies of BLQH systems at fractions of $\nu=2$. In 
particular, we think that our construction of coherent states on $\mathbb G_2$ will be relevant 
to study the interlayer macroscopic coherence in the BLQH system and a semiclassical study of quantum phase transitions, 
which is usually discussed in the simpler spin-frozen limit. 
Before, an interconnection between our CS and the usual variational wave functions of Laugling, Halperin and Jain 
\cite{Laughlin,Halperin,Jain,Jainbook} for correlated electrons
in the lowest Landau level would be in order. This is work in progress.

\section*{Acknowledgements}

Work partially supported by the
Spanish MICINN  and University of Granada under projects FIS2011-29813-C02-01 and PP2012-PI04, respectively.

\appendix
\section{Orthonormality of homogeneous polynomials\label{appendixortho}}

In order to prove the orthonormality relations
\be\la{}{}_{q'_a,q'_b}^{j',m'}|{}{}_{q_a,q_b}^{j,m}\ra=\int_{\mathbb G_2}
d\mu_\lambda(Z,Z^\dag){\varphi_{q'_a,q'_b}^{j',m'}(Z)}\overline{\varphi_{q_a,q_b}^{j,m}({Z})}=
\delta_{j,j'}\delta_{m,m'}\delta_{q_a,q'_a}\delta_{q_b,q'_b}\,,\label{orthorel}\ee
we shall adopt the following decomposition for
a matrix $Z\in\mathbb{G}_2$
\be Z=V_1\varXi V_2^\dag,\nn\ee
where
\be
V_u=\frac{1}{\sqrt{1+r_u^2}}\begin{pmatrix} 1 & r_u e^{i\alpha_u}\\ -r_u e^{-i\alpha_u} &1\end{pmatrix},\,
0\leq r_u<\infty, 0\leq \alpha_u< 2\pi, \, u=1,2,
\ee
are unitary matrices  and
\be \varXi=\begin{pmatrix} \rho_1 e^{i\theta_1} &0 \\ 0 &  \rho_2 e^{i\theta_2}  \end{pmatrix}, \,
0\leq \rho_u<\infty, 0\leq \theta_u< 2\pi, \, u=1,2,
\ee
Let us perform this change of variables to the invariant measure
(\ref{projintmeasure}). On the one
hand, the Lebesgue measure on $\mbC^4$ can be written as:
\be |dZ|=J(\rho_1,\rho_2)\rho_1 d\rho_1 d\theta_1 \rho_2 d\rho_2 d\theta_2 ds(V_1)ds(V_2),\nn\ee
with $ds(V_{u})=(1+r_u^2)^{-2}r_u dr_u d\alpha_u, u=1,2$, as in (\ref{haarmeasures2}), and
$J(\rho_1,\rho_2)=\frac{1}{2}(\rho_1^2-\rho_2^2)^2$ is the Jacobian determinant.
On the other hand, the weight factor in (\ref{projintmeasure}) adopts the form
\be\det(\sigma_0+Z^\dag Z)^{-\lambda-4}=
 ((1+\rho_1^2)(1+\rho_2^2))^{-\lambda-4}\equiv \Omega(\rho_1,\rho_2),\nn\ee
so that the invariant measure reads:
\be d\mu_\lambda(Z,Z^\dag)= c_\lambda
J(\rho_1,\rho_2)\Omega(\rho_1,\rho_2)\prod_{u=1}^2\rho_u d\rho_u d\theta_u  (1+r_u^2)^{-2}r_u dr_u
d\alpha_u.\nn\ee
where $c_\lambda\equiv \pi^{-4}(\lambda+1)(\lambda+2)^2(\lambda+3)$.

Let us denote by \[\cN_{j,m}\equiv\sqrt{\frac{2j+1}{\lambda+1}
\binom{\lambda+1}{2j+m+1}\binom{\lambda+1}{m}}\] the normalization constants of the
basis functions (\ref{basisfunc}). We want to evaluate:
\be\la{}{}_{q'_a,q'_b}^{j',m'}|{}{}_{q_a,q_b}^{j,m}\ra= \cN_{j,m}\cN_{j',m'} \int_{\mathbb{G}_2}d\mu_\lambda(Z,Z^\dag)
\overline{\det(Z)^{m}\cD^{j}_{q_a,q_b}(Z)}\det(Z)^{m'}\cD^{j'}_{q'_a,q'_b}(Z).
\ee
Using determinant properties, the Wigner's $\cD$-matrix
multiplication property
\be \sum_{q'=-j}^j
\cD^{j}_{qq'}(X)\cD^{j}_{q'q''}(Y)=\cD^{j}_{qq''}(XY)\label{Dmultprop}\ee
the transpositional symmetry
\be \cD^{j}_{qq'}(Y)=\cD^{j}_{q'q}(Y^T),\label{transposprop}\ee
and the fact that
$\det(V_{1,2})=1$ and that  $\varXi$ is diagonal, the previous
expression  can be restated as:
\bea&&\frac{\la{}{}_{q'_a,q'_b}^{j',m'}|{}{}_{q_a,q_b}^{j,m}\ra}{\cN_{j,m}\cN_{j',m'}}=
\sum_{q=-j}^j\sum_{q'=-j'}^{j'}c_\lambda\int_{\mbC^2}\prod_{u=1}^2\rho_u d\rho_u d\theta_u
J(\rho_1,\rho_2)\Omega(\rho_1,\rho_2) \nn\\ &\times&
\cD^{j}_{q,q}(\overline{\varXi})
\cD^{j'}_{q',q'}(\varXi)\det(\overline{\varXi})^m\det(\varXi)^{m'}\prod_{u=1}^2\int_{\mbS^2}
ds(V_u)\cD^{j}_{q_u,q}(\overline{V}_u) \cD^{j'}_{q'_u,q'}(V_u)\label{overlap-1}
\eea
Let us start evaluating the first integral.  For the diagonal matrix $\varXi$ we have that
$\cD^{j}_{q_a,q_b}(\varXi)=\delta_{q_a,q_b}(\rho_1 e^{i\theta_1})^{j+q_a}
(\rho_2 e^{i\theta_2})^{j-q_a}$, so that
\bea&& \cD^{j}_{q,q}(\overline{\varXi})
\cD^{j'}_{q',q'}(\varXi)\det(\overline{\varXi})^m\det(\varXi)^{m'}\nn\\
&&=\rho_1^{j+j'+q+q'+m+m'}\rho_2^{j+j'-q-q'+m+m'}e^{i(j'-j+q'-q+m'-m)\theta_1}e^{i(j'-j+q-q'+m'-m)\theta_2}.\eea
Integrating out angular variables gives the restrictions
\bea\int_0^{2\pi}\int_0^{2\pi}\cD^{j}_{q,q}(\overline{\varXi})
\cD^{j'}_{q',q'}(\varXi)\det(\overline{\varXi})^m\det(\varXi)^{m'}d\theta_1
d\theta_2\nn\\ =4\pi^2\delta_{q,q'}
\delta_{j+m,j'+m'}\rho_1^{2(j+q+m)}\rho_2^{2(j-q+m)}.\nn\eea
Integrating the radial part:
\bea 4\pi^2 c_\lambda\int_0^\infty \int_0^\infty
J(\rho_1,\rho_2)\Omega(\rho_1,\rho_2)\rho_1^{2(j+q+m)}\rho_2^{2(j-q+m)}
\rho_1d\rho_1\rho_2d\rho_2\nn\\ =
\frac{1+5q^2-(j+m)^2+(j+m+2q^2+1)\lambda}
{\pi^2(\lambda + 1)\tbinom{\lambda}{j+m+q}\tbinom{\lambda}{j+m-q}}\equiv {\cal R}_{j+m}^{q}\nn\eea
and putting all together in (\ref{overlap-1}) we have:
 \be \frac{\la{}{}_{q'_a,q'_b}^{j',m'}|{}{}_{q_a,q_b}^{j,m}\ra}{\cN_{j,m}\cN_{j',m'}}=
 \delta_{j+m,j'+m'}\sum_{q=-\min\{j,j'\}}^{\min\{j,j'\}}{\cal R}_{j+m}^{q}
\prod_{u=1}^2\int_{\mbS^2}
ds(V_u)\cD^{j}_{q_u,q}(\overline{V}_u) \cD^{j'}_{q'_u,q'}(V_u) \label{overlap-2}\ee
The last two integrals are easily computable. Actually they are a
particular case of the orthogonality properties of Wigner's
$\cD$-matrices. More explicitly:
\be\int_{\mbS^2} ds(V)\cD^{j}_{q_a,q_b}(\overline{V})
\cD^{j'}_{q'_a,q_b}(V)=\int_0^\infty\int_0^{2\pi}\frac{rdr d\alpha}{(1+r^2)^2}
\cD^{j}_{q_a,q_b}(\overline{V}) \cD^{j'}_{q'_a,q_b}(V)
=\delta_{j,j'}\delta_{q_a,q'_a} \frac{\pi}{2j+1}.\nn\ee
Going back to (\ref{overlap-2}) it results:
\be\la{}{}_{q'_a,q'_b}^{j',m'}|{}{}_{q_a,q_b}^{j,m}\ra=
\delta_{j,j'}\delta_{m,m'}\delta_{q_a,q'_a}\delta_{q_b,q'_b}(\frac{\cN_{j,m}}{2j+1})^2
\sum_{q=-j}^j \pi^2 {\cal R}_{j+m}^{q}.\nn\ee
Finally, taking into account the combinatorial identity:
\be \sum_{q=-j}^j \pi^2 {\cal R}_{j+m}^{q}=\frac{(2j+1)(\lambda+1)}{
\binom{\lambda+1}{2j+m+1}\binom{\lambda+1}{m}}\nn
\ee
and the explicit expression of the normalization constants $\cN_{j,m}$, we arrive at
the orthonormality relations (\ref{orthorel}).

\end{document}